\documentclass[preprint,12pt, a4paper]{elsarticle}

\usepackage{amssymb}
\usepackage{hyperref}
\usepackage{tabularx}
\usepackage{booktabs}
\usepackage{subcaption}

\usepackage{listings}
\usepackage{color}

\usepackage{caption}
\usepackage{graphicx}
\usepackage{rotating} 

\definecolor{dkgreen}{rgb}{0,0.6,0}
\definecolor{gray}{rgb}{0.5,0.5,0.5}
\definecolor{mauve}{rgb}{0.58,0,0.82}

\journal{SoftwareX}

\begin{document}
\renewcommand{\labelenumii}{\arabic{enumi}.\arabic{enumii}}

\begin{frontmatter}

\title{yProv4DV: Filling the Visualization Gap in Reproducible Research Workflows}

\author[]{G. Padovani}
\author[]{S. Fiore}
\address{University of Trento, Italy}

\begin{abstract}
While results visualization is a critical phase to the communication of new academic results, plots are frequently shared without the complete combination of code, input data, execution context and outputs required to independently reproduce the resulting figures. Existing reproducibility solutions tend to focus on computational pipelines or workflow management systems, not covering script-based visualization practices commonly used by researchers and practitioners. Additionally, the minimalist nature of current Python data visualization libraries tend to speed up the creation of images, disincentivizing users from spending time integrating additional tools into these short scripts. 
This paper proposes yProv4DV, a lightweight library designed to enable reproducible data visualization scripts through the use of provenance information, minimizing the necessity for code modifications. Through a single call, users can track inputs, outputs and source code files, enabling saving and full reproducibility of their data visualization software. 
As a result, this library fills a gap in reproducible research workflows by addressing the reproducibility of plots in scientific publications.
\end{abstract}

\begin{keyword}
Reproducibility \sep Provenance \sep yProv4DV \sep PROV-JSON \sep RO-Crate

\end{keyword}

\end{frontmatter}

\begin{table}[!ht]
\begin{tabular}{|l|p{6.5cm}|p{6.5cm}|}
\hline
\textbf{Nr.} & \textbf{Code metadata description} & \textbf{Please fill in this column} \\
\hline
C1 & Current code version & v1.0 \\
\hline
C2 & Permanent link to code/repository used for this code version & \url{https://github.com/HPCI-Lab/yProv4DV} \\
\hline
C3  & Permanent link to Reproducible Capsule & \\
\hline
C4 & Legal Code License   & GPLv3 \\
\hline
C5 & Code versioning system used & Git \\
\hline
C6 & Software code languages, tools, and services used & Python \\
\hline
C7 & Compilation requirements, operating environments \& dependencies & Prov \\
\hline
C8 & If available Link to developer documentation/manual & \url{https://hpci-lab.github.io/yProv4DV.github.io/} \\
\hline
C9 & Support email for questions & \textit{gabriele.padovani@unitn.it} \\
\hline
\end{tabular}
\caption{Code metadata}
\label{codeMetadata} 
\end{table}

\section{Motivation and significance}

Although the widespread use of Python and open source libraries improved the reproducibility and openness of data-driven research, reproducibility often still is a challenge, particularly in disciplines where data visualization of results plays a critical role~\cite{becker2019trackr}. While modern visualization libraries in Python (e.g., Matplotlib~\cite{Hunter:2007}, Seaborn~\cite{Waskom2021}, Plotly\footnote{https://plotly.com/}) make it straightforward to produce compelling graphical outputs, they also incentivize a codebase featuring numerous small scripts~\cite{perkel2018data}. Users may conduct several ad-hoc modifications to these files to produce consecutive image variations, without the ability to go back to a previous versions. Although tools such as Git\footnote{https://git-scm.com/}$^{,}$\footnote{https://wandb.ai/site} aid in code versioning aspects, or focus on reproducible Python environments\footnote{https://github.com/DavHau/mach-nix}, these programs are often so small that many modifications go untracked~\cite{liu2022promoting, isenberg2024state}. 
Existing efforts toward reproducible research have focused mainly on computational workflows, containerization, or version control of code and data~\cite{apostal2018containers, moreau2023containers}. However, these approaches overlook the specific needs of visualization scripts, which are frequently lightweight and detached from formal workflow management systems \cite{da2017characterization}. Researchers commonly produce visualizations in stand-alone scripts or notebooks, which are often modified to produce variants of the initial plot. As a consequence of this, even when input data, code and the environment specifics are archived, some images might not be reproducible, and the provenance of a figure, so how it was generated, from which inputs, under which parameters, and with which exact source code version, remains difficult to reconstruct. Provenance modeling standards, such as the W3C PROV data model~\cite{missier2013w3c, belhajjame2013prov}, provide a formal framework for representing the relationships between data, tasks and agents involved in computational processes. Despite their expressive power, these standards are rarely adopted in everyday data visualization practice due to the complexity of integration and the time required to add them into small scripts~\cite{herschel2017survey}. 
Therefore, there is a clear need for tools able to incentivize the use of formal provenance practices, improving its accessibility in small and simple workflows~\cite{silva2007provenance, ragan2015characterizing}. 

Many general purpose and targeted libraries for provenance collection have been proposed over the years, with the bulk of them being targeted at workflow management systems~\cite{davidson2008provenance}, at recently expanding fields such as machine learning (ML)~\cite{souza2022workflow, PADOVANI2025102298}, and at dockerized applications~\cite{kunstmann2024scientific}. 
Table \ref{tab:comparison} contains a comparison of the most popular libraries which are aimed to target reproducibility in small python scripts, and also includes information on what these tools collect and how can be integrated into existing code-bases. 

\begin{table}[h!]
    \centering
    \small
    \begin{tabular}{@{}lp{2.5cm}p{3.8cm}p{4cm}@{}}
        \toprule
        \textbf{Tool} & \textbf{Focus} & \textbf{Collection} & \textbf{Integration} \\ \midrule
        yProv4DV & Viz Reproducibility & Tracks filesystem calls to find dependencies. & In code: \texttt{start\_run()} \\ \midrule
        noWorkflow & Script Analysis and Lineage & Profiling and environment inspection. & Prepend to Run script: \texttt{now run script.py} \\ \midrule
        ReproZip & System Portability & Traces system calls to capture all files. & Prepend to run script: \texttt{reprozip trace python script.py}. \\ \midrule
        MLflow~\cite{zaharia2018accelerating} & ML Experiment Tracking & Automatic hooking into ML library APIs. & \texttt{mlflow.autolog()} in code \\ \midrule
        CDE~\cite{guo2018cde} & Linux Portability & Monitors Linux system calls during runtime. & Prefix command: \texttt{cde python script.py} \\ \bottomrule
    \end{tabular}
    \caption{Comparison of yProv4DV and similar provenance tracking libraries.}
    \label{tab:comparison}
\end{table}

To address the aforementioned gap, the yProv4DV (yProv for Data Visualization) library is proposed. yProv4DV is a lightweight Python utility which can automatically capture, package, and preserve the full reproducibility context of data visualization scripts. It is designed to minimize the number of directives that need to be invoked, and in most cases, simply importing the library and starting the run is sufficient. The system will record the source code, input files, generated outputs, and execution metadata of a visualization run, and organize these artifacts into a structured directory, producing a self-contained archive in the form of a RO-Crate~\cite{soiland2022packaging}, alongside W3C PROV–compliant provenance descriptions (JSON, DOT\footnote{https://graphviz.org/}, SVG). This enables interpretability through the PROV-JSON standard~\cite{niu2015interoperability}, as well as machine-readable traceability, facilitating reuse and long-term preservation of each script. 

Additionally, yProv4DV is one of the components provided in the yProv ecosystem~\cite{padovani2024software},  which has the aim of providing simple tools to collect and manage provenance information. The tools in the framework, such as yProv4ML~\cite{PADOVANI2025102298}, \cite{10.1145/3750720.3757295} which is designed to collect ML training information, and yProv4WFs~\cite{10825405}, which is integrated into several workflow management systems, are fully interoperable, and can facilitate the collection of multi-level provenance data by connecting files through Persistent Identifiers (PIDs) \cite{10386983}, \cite{MANZI2026108312}. 

In this paper, the design principles, architecture, and yProv4DV usage are presented, along with a demonstration of  how this framework supports reproducibility in visualization-centric scientific workflows.

\subsection{Software architecture}

Rather than introducing a separate workflow engine or requiring structural changes to code, yProv4DV operates as a lightweight runtime layer that observes, records, and packages the artifacts produced during execution.

\begin{figure}
    \centering
    \includegraphics[width=\linewidth]{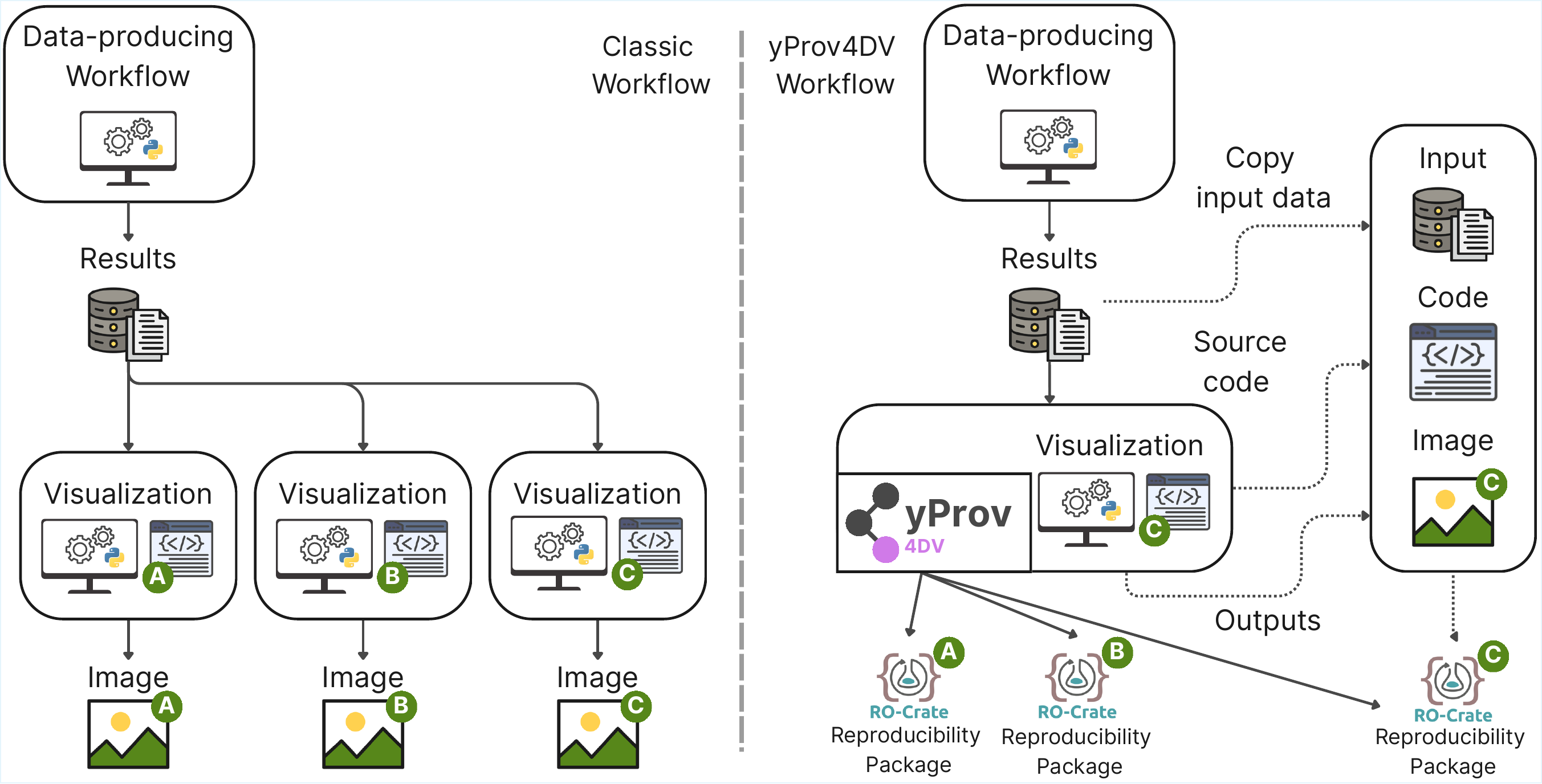}
    \caption{A standard data visualization pipeline compared to the one proposed when using yProv4DV. Currently, a set of Python scripts are created for the visualization of plots, often with a 1:1 ratio between images and scripts. With yProv4DV, a single Python script can be used, knowing that all runs generating an output will be tracked, and can be edited and re-used according to the user needs. }
    \label{fig:workflow}
\end{figure}

Figure \ref{fig:workflow} illustrates the difference between a classic visualization workflow and the one enabled by yProv4DV. In a conventional setting, individual Python scripts are typically developed to generate specific figures, resulting in a one-to-one relationship between program and output images. These source files often rely on external data and produce visual artifacts without preserving the full execution context in which they were generated. Without proper tracking of all files, this can lead over time to non-reproducible images, when scripts are modified.

The workflow proposed by yProv4DV restructures this process into even, depending on the user's preference, to a single visualization script. Each file's execution is tracked and packaged into a self-contained directory or RO-Crate, supporting reproducibility and facilitating the development process.

The yProv4DV architecture can be broadly subdivided into three main components: execution monitoring, artifact tracking, and provenance and reproducibility packaging.
The main idea behind the library is to create a runtime layer that intercepts key file operations during execution, without altering the user’s code logic. The module observes file reads performed by Python scripts, and monitors file creation events that occur as program runs. This tracking is fully transparent to the user, and can be customized to only observe specific types of files.

The artifact tracking component classifies all relevant resources involved in the visualization process into three broad categories:
 \begin{itemize}
     \item Inputs: which include files accessed during execution;
     \item Outputs: which are files generated by the script;
     \item Source code: which are all the Python scripts responsible for producing the visualization, including all the local modules imported by the program.
 \end{itemize}

These artifacts are copied into a dedicated provenance directory and organized into three respective sub-folders. 
Once artifacts are identified, and the execution has ended, the library constructs a standardized description of the execution using the W3C PROV format. It models the relationships between the main activity, which is the execution of the visualization script and all the entities (input files, source code, and output images), including an implicit agent, meaning the user or execution environment. This provenance description is then exported to JSON, image and to provenance graph formats, providing both machine-readable metadata and human-interpretable visual representations of the execution history. Finally, all data is packaged into a self-contained RO-Crate archive. 

Each execution of the script produces its own reproducibility package, meaning that multiple runs of the same script generate distinct provenance bundles. As shown in Figure \ref{fig:workflow}, the end goal of the library is to allow users to produce, from a single reusable visualization script, multiple fully documented outputs across several runs.

\subsection{Software functionalities}

As mentioned, yProv4DV is designed to be integrated easily into existing Python data visualization scripts, while automatically capturing the information necessary for reproducibility and provenance tracking. 
The only requirement from the user, is to import the library and call the \textit{start\_run} function:  

\begin{lstlisting}
import yprov4dv
yprov4dv.start_run(
    run_name : str = "experiment_run", 
    provenance_directory : str = "prov", 
    prefix : str = "yProv4DA", 
    default_namespace : str = "http://example.org/", 
    create_json_file : bool = False, 
    create_dot_file : bool = False, 
    create_svg_file : bool = False, 
    create_rocrate : bool = True,
    save_input_files : bool = True, 
    save_input_files_subset : bool = False,
    skip_files_larger_than : int = 50, 
    verbose : bool = False, 
)
\end{lstlisting}

This will start the tracking of newly generated files, as well as ingested inputs. Additionally, all libraries and modules imported before and after the calling of \textit{start\_run} will be included in the final RO-Crate. 

The behavior of the yProv4DV library can be customized modifying the parameters in the \textit{start\_run} function. For example, \textit{save\_input\_files} determines whether files tracked as input will be saved in the RO-Crate, and if \textit{save\_input\_files\_subset} is \textit{True}, then the library will only save the information used in the actual plot, and not the entire input file. 

As mentioned, the main feature of the library is the ability to automatically detect outputs produced during a script's execution. Any file generated by the script is automatically identified and registered as output in the provenance record. 
In practice, the library creates a set of snapshots of the work directory at both the start and at the end of the program execution, enabling the identification of newly created and modified files. All the detected files are then copied into the provenance directory to preserve the outputs of the visualizations, as well as the process required to reproduce them. In addition, yProv4DV provides automatic input tracking under a set of supported conditions. When a script reads a file — e.g. using Python's built-in \textit{open()} function — the library intercepts the operation and records the accessed file as an input in the provenance record.
This way the system is able to infer data dependencies directly from the script’s behavior without requiring users to manually declare which datasets or configuration files are being used, and especially without changing the user's execution script. 
To conclude the tracking, the library identifies all modules loaded by the initial script, which could be, for example, files written by the user to provide utility functions, and store them in a separate directory in the RO-Crate.  
All the files are then saved relative to their original location, which then allows for execution of the copied script without requiring additional actions from the user. 

yProv4DV supports a set of popular libraries which are often used for data visualization, such as pandas~\cite{mckinney2011pandas}, NumPy~\cite{harris2020array}, PyTorch~\cite{ketkar2021introduction}, Xarray\footnote{https://docs.xarray.dev/en/stable/index.html} and more. In case the user plot through an unsupported library, yProv4DV might not be able to fetch the input automatically. For these situations, the \textit{log\_input} and \textit{log\_output} directives have been exposed, which allow the user to inform the module on the files to track. 

\begin{lstlisting}
yprov4dv.log_input(path_to_untracked_input_file : str)
\end{lstlisting}

\begin{lstlisting}
yprov4dv.log_output(path_to_untracked_output_file : str)
\end{lstlisting}

In some cases, files used for visualization might be too large and copying them into an additional directory can be impractical and expensive. To address this, a file size threshold can be defined, allowing the library to ignore files exceeding a specified limit. This threshold can be configured through the \textit{start\_run} function, by setting the \textit{skip\_files\_larger\_than} parameter (in MB). 
In such cases, the reproducibility of the original script depends on whether the large input data can be shared through alternative means. To facilitate this the \textit{save\_input\_files\_subset} option can be set to \textit{True}, allowing only the subset of data actually used for plotting to be stored. The library is able to track all the major plotting functions from widely used libraries, such as pandas, matplotlib and seaborn, and to identify the specific dataframes involved in the visualization. This flag does not guarantee reproducibility, but rather it ensures that the subset of data required to reproduce the plotting logic  is included in the final RO-Crate.


\begin{lstlisting}
yprov4dv.untrack_file(path_to_tracked_output_file : str)
\end{lstlisting}

If a user wanted to prevent a specific file from being tracked, then the \textit{untrack\_file} directive can be used. Both input and output files that would normally be copied into the RO-Crate, when flagged, will be ignored. 

\begin{lstlisting}
yprov4dv.end_run()
\end{lstlisting}

Finally, the \textit{end\_run} function call has been defined to close the current execution and produce the whole reproducibility package. By default, the yProv4DV library calls this directive on exit, which makes it transparent to the user, however in some cases, such as when creating multiple charts in the same script, it allows each the visualization to be separated into several bundles.  

Table \ref{tab:performance_scaling} presents a performance benchmark based on 30 execution cycles, quantifying the library's overhead across increasing data scales. While the total execution time scales with the file size, the data reveals that the primary source of overhead is concentrated within the initialization and closing phases, and as the plot time increases, this wait time becomes more and more negligible.
\begin{table}[h]
\centering
\caption{Comparison of execution times (in seconds) and scaling factors between the yProv4DV version (With) and the baseline (Without) for files ranging from 8Kb to 8Mb.}
\label{tab:performance_scaling}
\begin{tabular}{lcccc}
\hline
\textbf{Metric} & \textbf{8Kb (s)} & \textbf{4Mb (s)} & \textbf{8Mb (s)} & \textbf{Scaling} \\ \hline
Init + Closing Time  & 1.9014 & 1.9725 & 1.9627 & 1.03x \\
Total Time (With)    & 2.6361 & 4.5090 & 5.9459 & 2.26x \\
Total Time (Without) & 1.1784 & 2.8126 & 4.1623 & 3.53x \\ \hline
\end{tabular}
\end{table}

\subsection{Sample code snippets analysis} 

The following code section shows how to integrate the yProv4DV library in an existing codebase: 

\begin{lstlisting}
# this can be imported even before the start_run function
from lib import elaborate

import yprov4dv
yprov4dv.start_run()

import pandas as pd
import matplotlib.pyplot as plt

# This is just to provide a reproducible script
import urllib.request
urllib.request.urlretrieve(
    "https://raw.githubusercontent.com/HPCI-Lab/" \
    "yProv4DV/main/assets/results.csv",
    "results.csv"
)

data = pd.read_csv("results.csv")
# In this case this is not necessary, 
# the file will be copied automatically
yprov4dv.log_input("results.csv")

data["second_series"] = elaborate(data["points"])

data.plot()
plt.savefig("example.png")

# In this case this is not necessary, 
# the file will be copied automatically
yprov4dv.log_output("example.png")

\end{lstlisting}

The only two necessary lines are the importing of the library and the \textit{start\_run} call, after which all opened and created files will be stored either as inputs or outputs. In the current example, the input file will be the one at \textit{"./results.csv"}, and the final image is \textit{"example.png"}. For explanatory purposes, the \textit{log\_input} and \textit{log\_output} directives are also shown, to give an example of how a user could decide to track additional files, but in this instance the calls would be redundant, as yProv4DV would track the files automatically.

%

Starting from the collected information, the user who produced this RO-Crate can simply call the following commands to reproduce the package. 

\begin{lstlisting}
# Run the original script, 
# creating the reproducibility package in "prov_0"
> python examples/main.py 
# Run the saved script, without moving from the original directory
> python prov_0/src/examples/main.py 
\end{lstlisting}

On the other hand, if the package is shared, the other user will have to set up the Python environment using the \textit{requirements.txt} file, and adjust the paths in the source file according to his file system.

\begin{lstlisting}
# Run the original script, 
# creating the reproducibility package in "prov_0"
> python examples/main.py 

# Adjust data paths (if the cwd is the same this step is not needed)
> vim prov_0/src/examples/main.py
# Install the requirements taken from the original env
> pip install -r prov_0/inputs/requirements.txt 

# Run the saved script, without moving from the original directory
> python prov_0/src/examples/main.py 
\end{lstlisting}


\section{Illustrative examples}

\begin{figure}
    \centering
    \includegraphics[width=\linewidth]{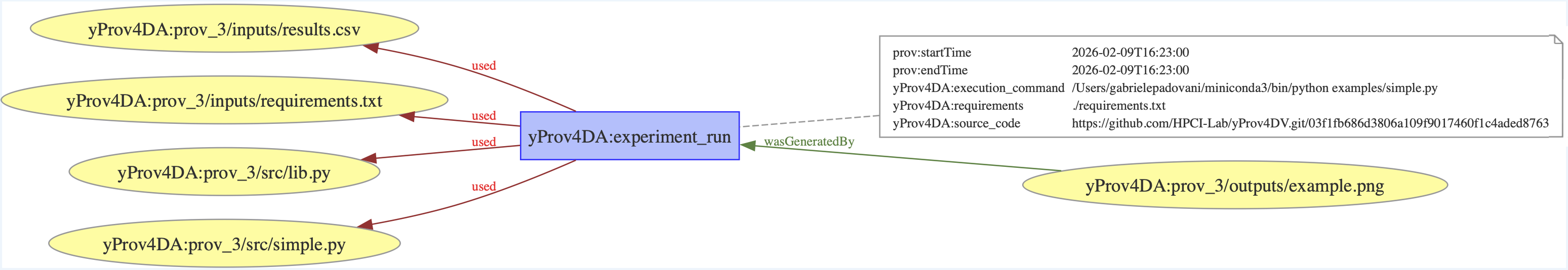}
    \caption{A brief example of a provenance graph containing the visualization pipeline is shown. On the left, inputs are shown, including the data used (\textit{results.csv}), the requirements for the Python environment and the modules and scripts used. On the right, the image output is shown. Attributes to the main activity include the Git hash to the source code, the execution command used and the start and end time. }
    \label{fig:experiment}
\end{figure}

Figure \ref{fig:experiment} presents a simplified graph generated from the execution of a visualization script to demonstrate how yProv4DV captures and represents provenance information. This example illustrates in practice how the library models the relationships between inputs, source code, execution metadata, and outputs according to the W3C PROV standard.

On the left side of the graph, the input entities used in the visualization process are shown. These include the datasets used to generate the figure (e.g., \textit{results.csv}), the \textit{requirements.txt} file describing the Python environment, and the source files (in this case \textit{lib.py}, \textit{simple.py}) that contribute to the execution. All these inputs are automatically detected when the program accesses them through standard file operations. Each of these entities is marked as "used" by the main activity and, in case the user explicitly logs the file through \textit{log\_input}, a parameter is specified to differentiate this possibility.

At the center of the graph is the activity node symbolizing the current run, so the visualization script (i.e., experiment\_run). This activity corresponds to a single run of the Python script tracked through yProv4DV. The activity is connected to an attribute node, containing execution metadata captured at runtime including, in this case:
\begin{itemize}
    \item the exact command used to run the script,
    \item the start and end timestamps of the execution,
    \item and the Git commit hash associated with the source code. 
\end{itemize}

On the right side of the graph, the output entities are shown (e.g., \textit{example.png}), and are marked as being "generated" by the execution activity. The same parameter distinguishing between user-logged and library-logged is also applied here, analogously to the inputs. 


An additional example of a use case of this library is provided through an European project consisting in the execution and gathering of information from small-scale HPC jobs into a dataset. Part of the data can be accessed at the AIMP-Bench repository\footnote{https://github.com/HPCI-Lab/AIMP-Bench}, which also includes a set of visualizations processed using yProv4DV. Starting from the elaboration of more than 300 ML trainings, an analysis phase was conducted to understand the behavior of different hardware and GPU configurations. A report\footnote{https://github.com/HPCI-Lab/AIMP-Bench/blob/main/AIMP.pdf} containing several charts has been produced, while all the reproducibility bundles created with yProv4DV have been uploaded in the \texttt{reprod} branch of the repository. 

The context of the repository shows a subset of the analysis, in particular for benchmarking the scaling capability of different GPUs starting from small ML use cases. 

\section{Impact}

yProv4DV has the potential to improve reproducibility practices in domains where data visualization is frequently used to explore results~\cite{isenberg2024state}. The ease of integration has the chance to lower the technical barrier to provenance capture, and hopefully incentivize researchers to move from ad hoc figure generation to systematically documented visualization workflows. Many users will not switch to large workflow management systems just to integrate provenance tracking, but obtaining reproducible packages with a single additional line might strive some to try.  

One of the key impacts of yProv4DV lies in making figure-level reproducibility feasible. While many reproducibility initiatives focus on entire computational pipelines, individual figures in publications often remain insufficiently documented. yProv4DV directly addresses this issue by packaging with a standardized format, each execution, its inputs, source code, and execution context. This capability enhances the integrity of scientific reporting and simplifies peer review, replication of studies, and educational reuse of published materials. The integration of W3C PROV–compliant metadata and RO-Crate packaging further improves the usability of yProv4DV, especially when users have to share their own work. The generated artifacts are interoperable with existing provenance analysis tools, lowering the barrier to the creation of standardized archives. As research infrastructures increasingly emphasize FAIR (Findable, Accessible, Interoperable, Reusable) principles~\cite{wilkinson2016fair}, yProv4DV, along with the yProv ecosystem, will help contribute to making visualization scripts more FAIR-compliant, especially if each artifact can be easily uploaded with its own associated PID.

In conclusion, yProv4DV promotes Open Science principles through its provenance aware capabilities, which ensure the reproducibility of visualization workflows in scientific research and publications, without imposing additional burdens on users. Over time, we hope its integration will be able to shift the common practices toward such a provenance-aware model.

\section{Conclusions and Future Work}
In this paper, we introduced yProv4DV, a lightweight Python utility designed to address a critical gap in reproducible research: the lack of systematic provenance capture for data visualization scripts, like those needed for scientific papers. 
The library is designed to track source code, inputs, outputs, and execution metadata, packaging them into W3C PROV–compliant JSON files and RO-Crates. yProv4DV provides a standard solution for Python scripts' reproducibility. Its minimal integration requirements and customization make it accessible to a wide range of users, and aims to expand the yProv ecosystem of tools~\cite{padovani2024software} that improve the production and handling of provenance files, integrating it into visualization scripts without the need for complex workflow orchestration tools. 


Future work related to this library will include: integration with more visualization libraries, allowing for more automatic reproducibility with a path tracing system, and a thorough validation of this library's functionality on larger use cases, in addition to the one already present for the AIMP Benchmark. Additionally, with the improvement of ML-based visualization tools~\cite{zhu2026paperbanana}, tools such as yProv4DV will have to evolve to keep pace with new and increasingly advanced scenarios: further work will explore interoperability with data repositories, compatibility with other existing provenance standards and data management platforms, such as mlflow servers, enabling direct publication of provenance artifacts. Finally, further evaluations to assess the effectiveness of yProv4DV in improving reproducibility practices across different research domains will be addressed.

\section*{Acknowledgements}

\textit{This work was partially funded under the National Recovery and
Resilience Plan (NRRP), Mission 4 Component 2 Investment 1.4 - Call for tender No. 1031 of 17/06/2022 of Italian Ministry for University and Research funded by the European Union – NextGenerationEU (proj. nr. CN\_00000013), and the RI-SCALE project (Grant Agreement 101188168)}

\bibliographystyle{IEEEtran}
\bibliography{refs}

\end{document}